# Optical forces, torques and force densities calculated at a microscopic level using a self-consistent hydrodynamics method

Kun Ding and C. T. Chan[†]

*Department of Physics and Institute for Advanced Study, The Hong Kong University of Science and Technology, Clear Water Bay, Hong Kong*

[†] Corresponding E-mail: phchan@ust.hk

**Abstract**

The calculation of optical force density distribution within a material is challenging at the nanoscale, where quantum and non-local effects emerge and macroscopic parameters such as permittivity become ill-defined. We demonstrate that the microscopic optical force density of nanoplasmonic systems can be defined and calculated using a self-consistent hydrodynamics model that includes quantum, non-local and retardation effects. We demonstrate this technique by calculating the microscopic optical force density distributions and the optical binding force induced by external light on nanoplasmonic dimers. We discover that an uneven distribution of optical force density can lead to a spinning torque acting on individual particles.

**PhySH:**

Within the framework of classical electrodynamics, optical forces acting on an object can be obtained by integrating electromagnetic stress tensors over a boundary enclosing the object [1,2]. There are different formulations of macroscopic electromagnetic stress tensors [2-6], each of which gives exactly the same result if the boundary is in a vacuum but different results if the boundary cuts across a material [7-14]. If we go one step further to find the distribution of force (force density) within a material, different electromagnetic tensors also yield different results. Which tensor is the correct one is a very complicated issue [8-18]. Calculating the force and force density in nanosystems is even more difficult when quantum effects such as electron "spill-out", non-locality and charge tunneling occur [19-24]. Classical stress tensors that require macroscopic permittivity and permeability simply cannot be used.

An ab-initio approach that determines microscopic charge/current and fields should be ideal for obtaining optical force densities at the microscopic level for nanosystems. However, most ab-initio algorithms do not include the retardation effects of external electromagnetic (EM) waves, which can be a problem because optical scattering forces are dominated by retardation. While it is possible to combine ab-initio density functional methods with Maxwell equations [25,26], computation is extremely demanding. In this study, we use a self-consistent hydrodynamics model (SCHDM) [27-30] to calculate microscopic optical force densities in plasmonic systems. This method solves Maxwell equations and the equation of motion for electrons on an equal footing and can hence provide a description of optical force densities at the microscopic level, taking quantum and retardation effects into account at a reasonable computation cost.

We demonstrate the method by investigating light-induced forces and the microscopic optical force densities of metallic nanoparticle dimers illuminated by an external light source. The microscopic optical force densities also give the optical torque acting on each nanoparticle, and we find a notable spinning torque when the gap size is at the nanometer scale. The optical binding force and spinning torque are closely correlated with the evolution of plasmonic modes as the particles approach each other. We consider two-dimensional (2D) configurations in which the particles are cylinders and the k-vector of the external EM field is normal to the axis of the cylinders. We use the standard jellium model, which offers an adequate description for simple metals, and SCHDM is known to be suitable for simple metals [27-30]. We consider sodium particles as our prototype, with ion density $n_{ion}$ defined as $n_{ion} = \frac{3}{4\pi (r_s a_H)^3}$, where $a_H = 0.529 \text{Å}$ is the Bohr radius and the dimensionless quantity $r_s = 4$.

Let us start with the dimer configuration shown in the inset of Fig. 1(b). Two identical plasmonic circular cylinders are placed close to each other, separated by a gap of size $d_{gap}$. The yellow regions represent the positively-charged jellium background with radii $r_a$. The first step of SCHDM involves determining the electronic ground state that minimizes a density functional subjected to constraints (chemical potential and electron number), which requires the numerical calculation of the equilibrium electron density $n_0$ and effective single electron potential $V_{eff}$ [31,32]. Once the ground states are obtained, the excited state calculations can be performed numerically [31,32] by coupling the linearized equations of motion for the electron gas with Maxwell equations, implying that retardation is automatically included. The induced charge density $\rho_1$, current density $\mathbf{J}_1$, and microscopic electric (magnetic) fields $\mathbf{E}_1 (\mathbf{B}_1)$ are solved numerically using the ground state results from the previous step. The incident light propagates in the y-direction ($\mathbf{k}_{inc} \parallel \hat{y}$), and the electric fields are polarized in the x-direction, i.e. $\mathbf{E}_{inc} = \hat{x} E_0 e^{i\mathbf{k}_{inc} \cdot \mathbf{r}}$ (see Supplemental Material, Section I, for numerical details [33]).

The calculated absorption cross section as a function of frequency for the dimer for different gap sizes are shown in Fig. 1(a). For comparison, we calculate the absorption spectrum of a single cylinder by setting $d_{gap} = \infty$, as shown by the panel marked by $\infty$ in Fig. 1(a). The single cylinder spectrum has two peaks. The lower-frequency main peak is the dipole plasmon resonance, which is red-shifted from the classical resonance frequency $4.165\text{eV} \left( = \omega_p / \sqrt{2} \right)$ due to the electron spill-out effect [34-36]. The higher-frequency minor peak is the Bennett plasmon mode (denoted by *M*), which can only appear in quantum models that can handle electron spill-out effects [37-39]. When $d_{gap} = 4.0\text{nm}$, two well-separated absorption peaks remain in the spectrum, traceable to those of a single cylinder [29]. Reducing the gap size splits the main absorption peak into two, as is exemplified by the spectrum at $d_{gap} = 1.0\text{nm}$ in Fig. 1(a). The lower-frequency peak is the dipole mode (marked by *D*), as the induced electric dipoles of the two cylinders are in phase. The higher-frequency peak is a quadrupole mode (denoted by *Q*). The split between the D and Q modes increases as the gap size decreases. When the gap is small enough to allow the tunneling of electrons through it, charge transfer plasmonic modes (denoted by *C1* and *C2*) emerge [40,41]. These charge transfer modes are consistent with ab-initio [40-46] and experimental results [47-52] observed in similar systems, indicating that our SCHDM calculations capture the essential physics of such systems (see Supplemental Material, Section II, for mode profiles [33]).

We now come to the force calculations. Optical forces are usually calculated using various forms of electromagnetic stress tensors, as specified by the macroscopic local permittivity $\varepsilon$ and permeability $\mu$ of the material. In the nanoscale, when non-locality and quantum effects become important, local constitutive parameters become meaningless. Our method can determine the microscopic electric and magnetic fields, allowing us to calculate the light-induced optical force and torque without using macroscopic values of $\varepsilon$ and $\mu$. Microscopic induced charges and currents are available point by point and can be used to calculate the force density. Within the SCHDM framework, the time-averaged optical forces due to a time-harmonic external field can be calculated by examining the time-averaged Lorentz force densities of the electron gas, defined as $\mathbf{f} = \langle \rho_1 \mathbf{E}_1 + \mathbf{J}_1 \times \mathbf{B}_1 \rangle$, in which $\mathbf{E}_1$ and $\mathbf{B}_1$ are microscopic fields including both incident and scattering fields. The force density distributions and the integrated total forces are shown in Fig. 2. As with the configuration shown in the inset of Fig. 1, the light is propagated in the y-direction with electric field polarized along the x-direction. Integrating the y-component of the time-averaged Lorentz force densities in the entire calculation domain ($\mathbf{F}_y = \int_{\Omega_T} \mathbf{f}_y \, d\mathbf{r}$) gives the total optical forces acting on the system in the direction of the incident light, as shown by the solid blue lines in Fig. 2(a) for $d_{gap} = 0.1\text{nm}$. Each peak in the total optical force spectrum can be identified with a corresponding absorption peak and labeled accordingly, as shown in Fig. 1(a) as the particle gains mechanical momentum after absorbing incident photons. For comparison, we also calculate the total optical forces by finding the surface integral of the time-averaged Maxwell stress tensor $\langle \mathbf{T} \rangle$ in the far field, as shown by the red open circles in Fig. 2(a). The results calculated using the Lorentz force formula and the Maxwell tensor are the same because we are using microscopic fields.

We plot the force distribution of $\mathbf{f}_y$ for the charge transfer modes C1 and C2 in Figs. 2(b) and 2(c) respectively when $d_{gap} = 0.1\text{nm}$. We observe that the optical force density is concentrated in a very small region along the boundary and the y-component of the optical force density for the dimer is at a maximum near the vacuum gaps but not at the smallest gap position when tunneling occurs. The sign of the force density changes near the gap (red/blue color in Fig. 2), indicating that the surfaces near the gap experience large stresses. The magnitude of the optical force density for the C2 mode is one order larger than that of the C1 mode, consistent with the total force results in Fig. 2(a).

The distribution of $\mathbf{f}_x$ for these charge transfer modes is shown in Figs. 2(e) and 2(f) respectively. Similarly to the y-component, the x-component of the force density is concentrated in a small boundary region but is one order of magnitude larger than $\mathbf{f}_y$ although the incident light momentum is in the y-direction. The optical binding forces between the plasmonic particles can be obtained using classical electrodynamics if they are well separated but not when the particles are so close that nonlocality is important or when tunneling occurs. Here, we can obtain the optical binding forces by calculating the volume integral of the time-averaged Lorentz force density in the left (or right) particle calculation domain $\Omega_L$ ($\Omega_R$) as $\mathbf{F}_x = \int_{\Omega_L} \mathbf{f}_x \, d\mathbf{r}$. The domain $\Omega_L$ and $\Omega_R$ is chosen to be symmetric about the y-axis, and the boundaries of these domains are shown by the dashed lines in Fig. 1(b). The calculated optical binding force spectra as a function of various gap sizes are shown in Fig. 1(b). As the magnitude of light-induced forces depend on external light intensity, we normalize the calculated force with respect to $F_0$, which is the optical force of a perfect absorber of the same geometric cross section as a single cylinder. We find a significant attractive force acting on each particle and hence a light-induced binding force between these two particles. For comparison, we also calculate the surface integral of time-averaged Maxwell stress tensor $\langle \mathbf{T} \rangle$ on the boundary of the domain $\Omega_L$, namely $\int_{\partial \Omega_L} \langle \mathbf{T} \rangle \cdot \hat{n} \, dS$. The results are plotted by the red open circles in Fig. 2(d) for $d_{gap} = 0.1 \text{nm}$. Comparing Figs. 2(a) and 2(d), we see that the forward scattering force (y-direction) can be slightly enhanced by resonance, which can also be seen from the absorption cross section in Fig. 1(a), but the optical binding forces (x-direction) are significantly amplified (by hundreds of times the value of $F_0$) although it is in the transverse direction. These excellent agreements demonstrate that if the microscopic charges, currents, and fields can be obtained, then a Maxwell stress tensor is the only suitable choice for calculating the surface integral.

Comparing Fig. 1(a) with Fig. 1(b), we find that not all absorption resonance gives rise to optical binding. Strong binding forces are found for the mode D (which becomes C2 when tunneling occurs). We plot the magnitudes of these maximum binding forces for various values of $d_{gap}$ in Fig. 3(a). The binding force increases rapidly as the gap size decreases, reaching a maximum at $d_{gap} \cong 0.3 \text{nm}$. When the gap size further decreases, the binding force drops due to the charge transfer in the gap. If we naively assume that the electromagnetic energy stored in the dimer is dominated by one mode, the energy can be written as $U = N\hbar\omega$, where $N$ is the number of photons in the resonance mode. The binding forces can then be written as $F = -\frac{\partial U}{\partial x} = -N\hbar \frac{\partial \omega}{\partial x}$. Consider the ratio as $\mathbf{F}_x / \left( \frac{\partial \omega_{D,C2}}{\partial d_{gap}} \right)$, which is plotted in Fig. 3(b). A constant ratio indicates that the character of this mode remains the same; otherwise the mode must change to other modes. When $d_{gap} > 0.5 \text{nm}$, the ratio is nearly constant. A sharp transition occurs at $d_{gap} < 0.5$nm, which can be used as a marker to characterize the transition between the dipole (D mode) and charge transfer mode (C2). The binding forces of the C1 mode are also substantial for small gap sizes. These results suggest that the optical binding forces offer a useful indicator of changes in plasmonic modes as system configuration changes that may otherwise be much more tedious to trace (such as by examining mode profiles).

The microscopic optical force densities can also be used to qualitatively predict the behavior of the system. For example, visual examination of Figs. 2(b) and 2(c) indicates that there must be a light-induced spinning torque acting on each cylinder as the force densities in each cylinder are not symmetric about their own center. We calculate the optical torque of the right cylinder by calculating the volume integral $\boldsymbol{\tau} = \int_{\Omega_R} (\mathbf{r} - \mathbf{r}_c) \times \mathbf{f} \, d\mathbf{r}$, where $\mathbf{r}_c$ is the center of the right cylinder. Figure 3(c) shows the z-component of the optical torque for $d_{gap} = 4.0$nm, $d_{gap} = 1.0$nm, and $d_{gap} = 0.1$nm respectively. We observe that the torque is negative (rotating clockwise for the right cylinder) for the charge transfer plasmonic modes and the D mode and positive for the Q mode. This is because in the C1, C2, and D modes, the optical force densities are higher near the gap regions, and the force direction is in the general direction of light propagation (red arrows), and so the right/left cylinder rotates in the clockwise/counter-clockwise direction, as shown in the left inset of Fig. 3(c). On the contrary, the optical force densities for the Q mode are lower near the gap regions, and hence the right/left cylinder rotates in the counter-clockwise/clockwise directions. These torques are significant at the nanoscale. We estimate that the angular acceleration gained by the single cylinder is approximately $10^{14}$ rad/s² if the incident light power is 1.0 mW/µm².

To demonstrate the versatility of the method, we perform similar calculations for two sodium triangles (Fig. 4). The calculated absorption spectra for different gap sizes are shown in Fig. 4(a). When the gap size is large, the absorption spectrum is similar to that of a single triangle, with three major absorption peaks in the spectrum (two edge modes and one face mode [31]). With a decreasing $d_{gap}$, the fundamental edge mode (Ed$_1$) transitions to the charge transfer mode C2, and another charge transfer mode C1 appears. We show the x-component of optical force density for the modes C1 and C2 in Figs. 4(e) and 4(f) respectively. The maximum force density also occurs near the vacuum gap, similarly to the cylinder dimer. The maximum binding force occurs in the modes Ed$_1$ and C2. To illustrate the mode transitions, we also plot the optical binding forces between these two triangles for different gap sizes in Fig. 4(b) and plot the maximum binding force as a function of $d_{gap}$ in Fig. 4(c). The ratio $\mathbf{F}_x / \left( \frac{\partial \omega_{Ed1,C2}}{\partial d_{gap}} \right)$ is plotted in Fig. 4(d), which shows that the mode transition from Ed$_1$ to C2 occurs at $d_{gap} \cong 0.4 nm$. The binding forces in the C1 mode are also greater than in the C2 mode. As a result, the binding forces can still quantitatively predict the mode transitions in the bow tie structures, similarly to the two-cylinder dimer configuration. In addition, the spinning torque for each triangle induced by external light is also nonzero but one order of magnitude smaller than that of the cylinder (see Supplemental Material, Section III [33]).

We note that the Lorentz force is the electromagnetic force term that depends linearly on external fields. There are other internal force terms, such as those of the electron fluid due to changes in the local chemical potential. In addition, there are chemical bonding forces which in the language of DFT are called Hellmann–Feynman forces that are determined by the electronic eigenstates, which are weakly dependent on the external field. These quantum forces do not depend explicitly on external fields and are not included in our electromagnetic force consideration. They should be calculated using DFT if desired. All of these forces have a short range and exist even in the absence of external EM waves, but the optical binding forces here are caused explicitly by the scattering of EM waves between these two particles and are therefore fairly long-range forces, as shown in Fig. 3(a).

In summary, we have calculated the microscopic optical force densities for dimerized nanoplasmonic particles using a self-consistent hydrodynamic model. We show that the microscopic optical force density and binding forces can be defined and calculated although quantum effects and non-locality are significant. Furthermore, the binding force spectrum provides a useful method of tracing plasmonic mode evolution. We also show that the uneven distribution of optical force density can lead to a strong spinning torque acting on each nanoparticle.

**Acknowledgements**

This project was supported by the Hong Kong Research Grants Council (grant no. AoE/P-02/12).

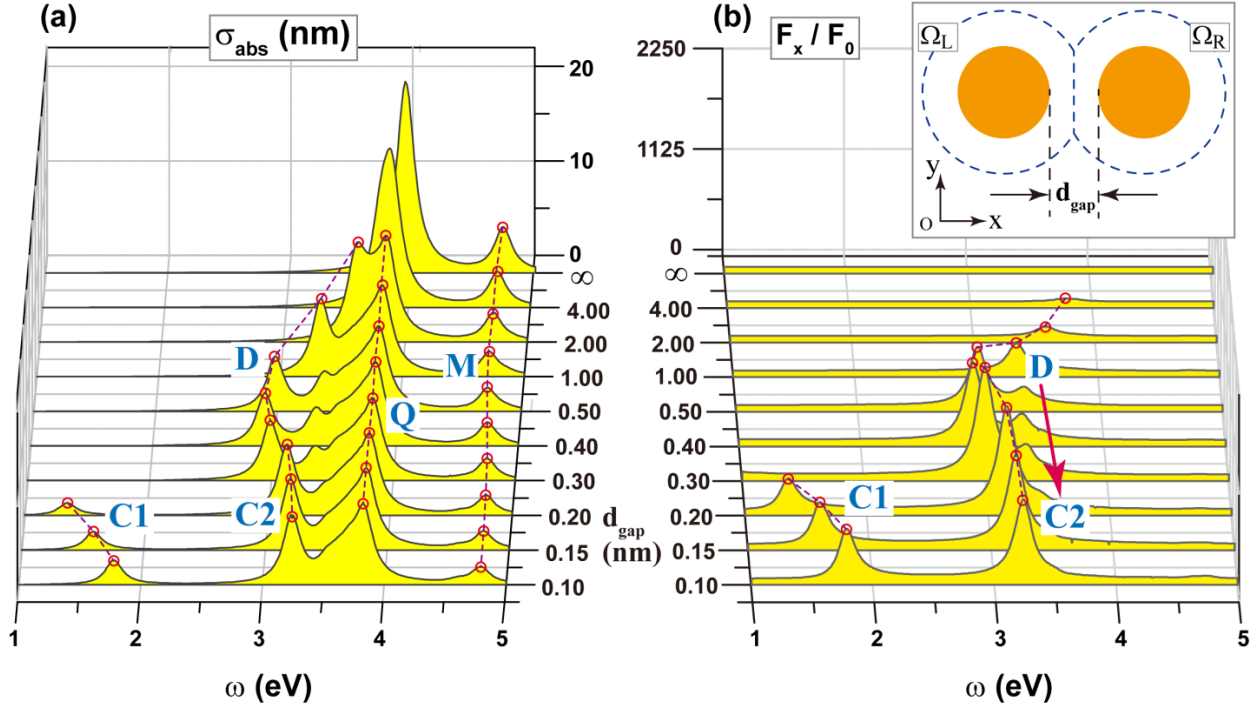

Figure 1. (a) Absorption cross section $\sigma_{abs}$ and (b) optical binding force $\mathbf{F}_x$ for different separations ($d_{gap}$) of cylinder dimers under plane wave illumination with $\mathbf{k}_{inc} \parallel y$, $\mathbf{E}_{inc} \parallel x$ and $E_0 = 1.0$ V/m. $F_0$ is the optical force for a perfect absorber of the same geometric cross section as a single cylinder. The dashed lines and open red circles label different plasmonic modes. The inset in (b) shows a plasmonic cylinder dimer with the yellow region marking the jellium background, and $\Omega_L$ and $\Omega_R$ are the calculation domains for the left and right particles respectively.

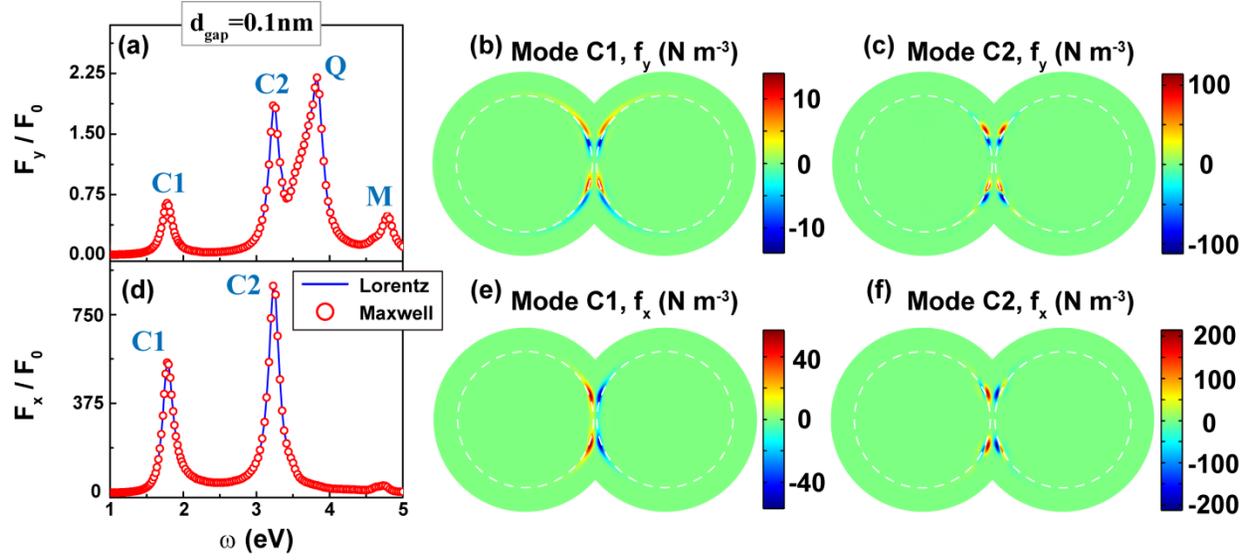

Figure 2. (a) Total force in the y-direction and (d) optical binding forces (x-direction) for $d_{gap} = 0.1$ nm. Note the agreement between results obtained by integrating Lorentz force density (solid blue lines) and Maxwell stress tensor (red open circles). (b) and (c) show the y-components of Lorentz force densities for modes C1 and C2 respectively. (e) and (f) show the x-components. The white dashed lines mark the jellium boundaries.

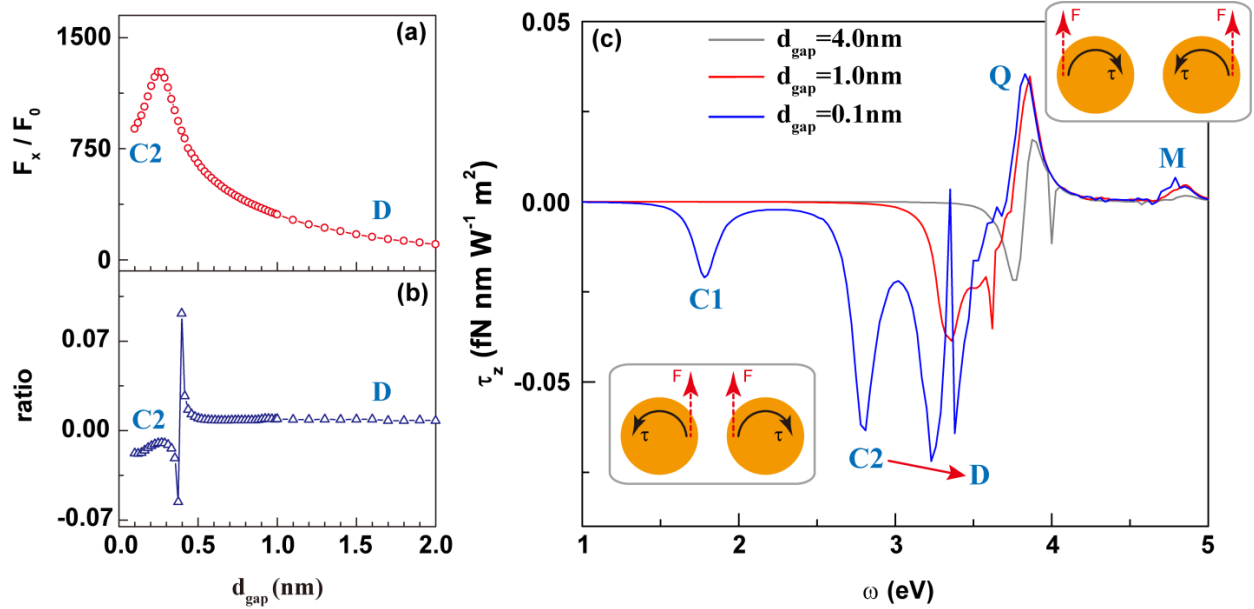

Figure 3. (a) The maximum optical binding forces as a function of gap sizes. (b) The ratio, defined as $\mathbf{F}_x / (\partial \omega_{D,C2} / \partial d_{dgap})$, as a function of gap size. (c) Optical torque (z-component, normalized to per unit length) acting on the cylinder on the right for $d_{gap} = 4.0\text{nm}, 1.0\text{nm}, 0.1\text{nm}$. The insets show that the location of maximum force densities (red dotted arrow) determines the sense of rotation (black arrow).

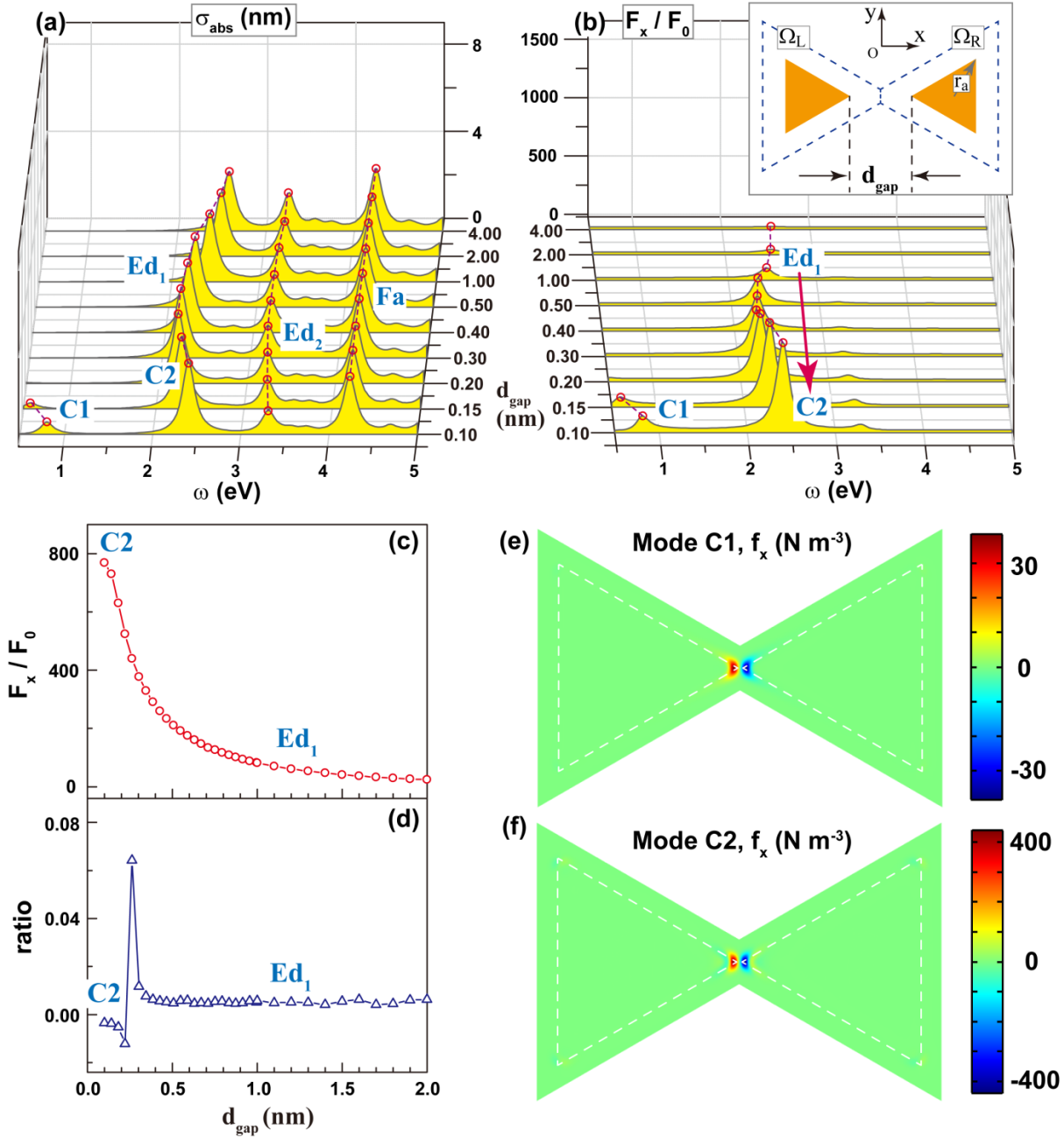

Figure 4. (a) Absorption cross section $\sigma_{abs}$ and (b) optical binding force for decreasing gap sizes of the two-triangle dimer (illustrated in the inset) under single plane wave illumination with $\mathbf{k}_{inc} \parallel y$, $\mathbf{E}_{inc} \parallel x$ and $E_0 = 1.0$ V/m. The dashed lines and open circles mark different plasmonic modes. (c) The maximum optical binding force as a function of gap size. (d) The ratio, defined as $\mathbf{F}_x / (\partial \omega_{Ed1,C2} / \partial d_{gap})$, as a function of gap size. The x-component of Lorentz force density for the modes C1 and C2 are plotted in (e) and (f) respectively.

# Supplemental Material --- Optical forces, torques and force densities calculated at a microscopic level using a self-consistent hydrodynamics method

Kun Ding and C. T. Chan[†]

*Department of Physics and Institute for Advanced Study,
The Hong Kong University of Science and Technology, Hong Kong*

[†] Corresponding E-mail: phchan@ust.hk

## I. Detailed numerical results

Figure S1(a) shows the two identical nano plasmonic circular cylinders, each with radius $r_a$. They are placed close to each other, separated by a gap distance $d_{gap}$. The yellow regions stand for the positive charged jellium background with radius $r_a$. The calculated $(n_0 - n_j)/n_{ion}$ of the two sodium circular cylinders with $r_a = 2.0$nm and $d_{gap} = 0.1$nm are shown in Fig. S1(c). The $n_j$ denotes the jellium density and $n_0$ denotes the equilibrium electron density. We see that the electron density inside the gap is nonzero because the tunneling effect of electrons (see also Fig. S2). The values of the effective potential inside the gap are much smaller than that in the vacuum outside the cylinders (Fig. S2), indicating that the electrons are much easier to tunnel through the gap than the vacuum.

Figure S1(b) shows the two identical nano plasmonic triangles with circumradius $r_a$ are placed close to each other, separated by a gap distance $d_{gap}$. The calculated $(n_0 - n_j)/n_{ion}$ of the bow tie structure with $r_a = 2.0$nm and $d_{gap} = 0.1$nm are shown in Fig. S1(d). The electron density inside the gap is also nonzero because the tunneling effect of electrons. The values of the effective potential inside the gap are much smaller than that in the vacuum outside the triangles (Fig. S2), indicating that the electrons are much easier to tunnel through the gap than the vacuum.

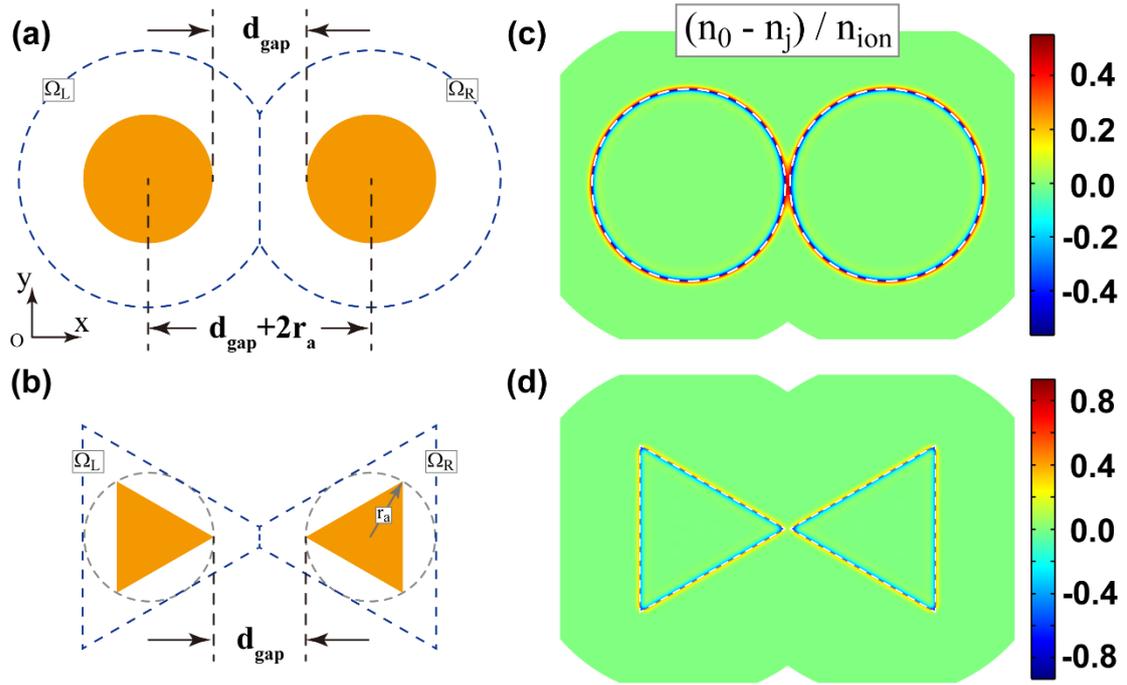

**Figure S1.** Schematic picture of the plasmonic dimers (a) circular cylinders and (b) triangular cylinders. Yellow region stands for the jellium background with circumradius $r_a$, $\Omega_L$ ($\Omega_R$) is the calculation domain for the left-sided (right-sided) particle. Calculated electron distribution $(n_0 - n_j)/n_{ion}$ in the ground state for the two-cylinder case and the two-triangle case with circumradius $r_a = 2$nm and $d_{gap} = 0.1$nm are shown in (c) and (d), respectively.

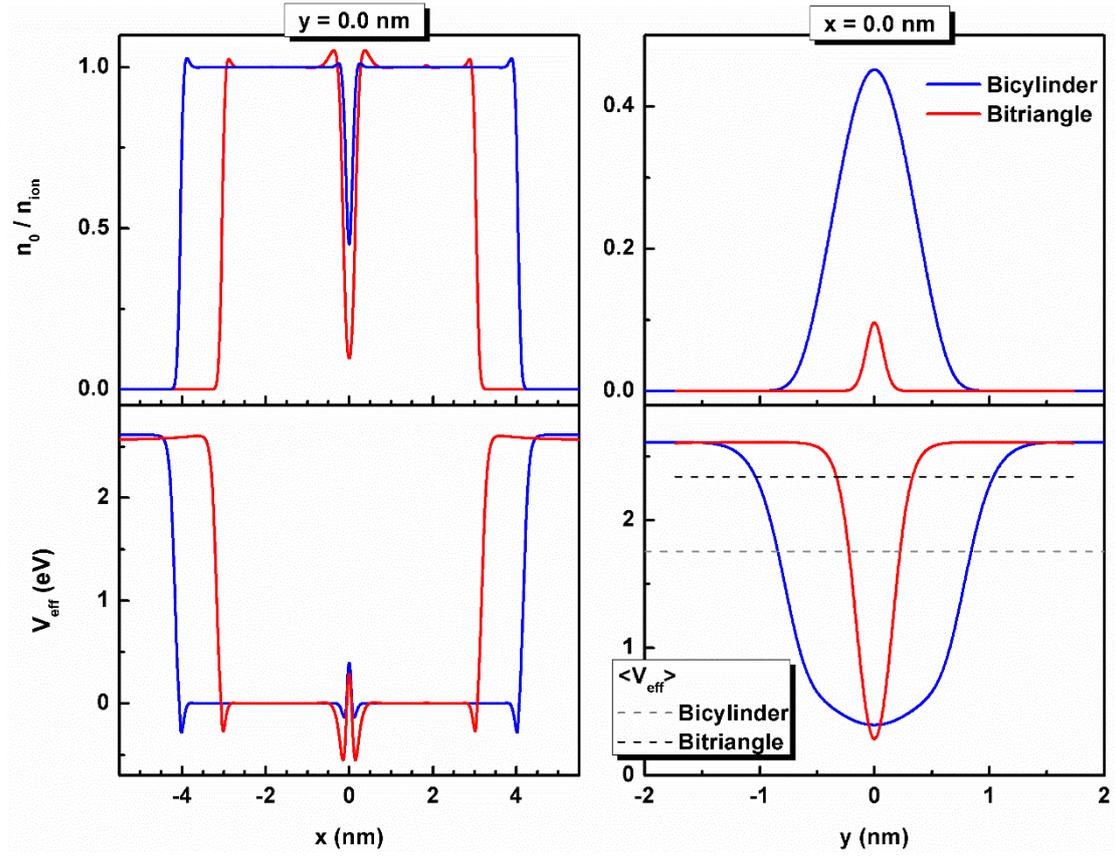

**Figure S2.** Electron distributions $n_0/n_{ion}$ and effective potential $V_{eff}$ for the two-cylinder dimer (blue lines) and the two-triangle dimer (red lines). The left panels show the distributions as a function of x for y=0nm, and the right two panels show the distributions as a function of y for x=0nm.

After the ground states are obtained, we need to carry out the excited state calculations in order to study the absorption and scattering properties of the nanoparticle systems. Figure S3 shows the calculation domain of the excited states (denoted by $\Omega_T$), which is chosen to be a large square with the dimerized nanoparticle in the center, and the boundary conditions of this region are perfect matched layer (PML).

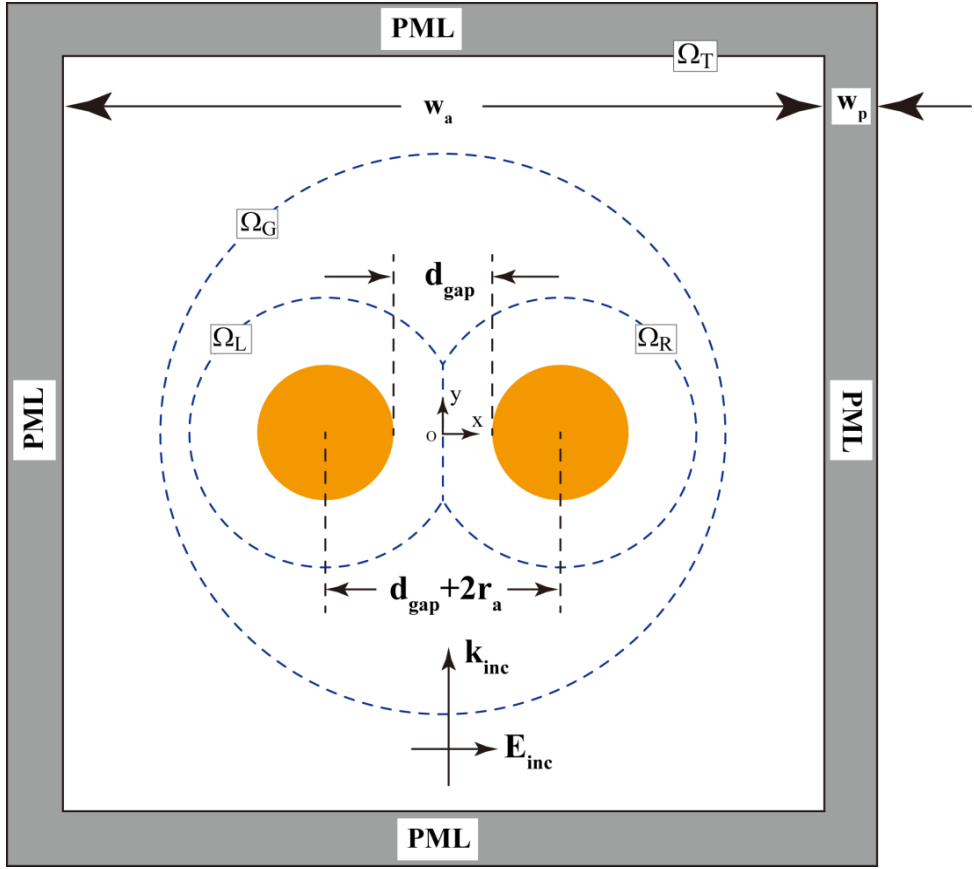

**Figure S3.** Schematic picture of the calculation domain and the boundaries. The dashed lines stand for the virtual boundaries and the surrounding boundaries of the whole domain are perfect matched layer (PML). Throughout this work, we set the radius of $\Omega_G$ to 8.0nm, the width $w_a$ to 400nm, and the thickness of PML layer $w_p$ to 40nm.

## II. Mode profiles of the two-cylinder dimer

The calculated absorption cross sections as a function of frequency for the two-cylinder system for different gap sizes are shown in Fig. S4(a). For comparison, we calculate the absorption spectrum of a single cylinder by setting $d_{gap} = \infty$, as shown by the panel marked by $\infty$ in Fig. S4(a). The lower-frequency main peak is the dipole plasmon resonance, which is red-shifted compared to the classical resonance frequency $4.165\text{eV}\left(=\omega_p/\sqrt{2}\right)$. The minor peak at a higher frequency is the Bennett plasmon mode (denoted by *M*), which can only appear in quantum models that can handle electron spill-out effects. When $d_{gap} = 4.0\text{nm}$, there are still two well separated absorption peaks in the spectrum, traceable to those of a single cylinder. This indicates that the coupling between these two particles is rather weak at that distance. Decreasing the gap size splits the main absorption peak into two peaks. See for example the spectrum at $d_{gap} = 1.0\text{nm}$ in Fig. S4(a). The lower frequency peak is the dipole mode (denoted by *D*), as the induced electric dipoles of the two cylinders are in phase, as shown by the induced charges $\rho_1$ and currents $\mathbf{J}_1$ at some particular time of this mode in Fig. S4(b). The higher frequency peak is a quadrupole mode (denoted by *Q*), and the character of this mode could be seen from the charge and current distributions plotted in Fig. S4(c). The splitting between the D and Q mode increases as the gap size decreases.

When the gap size is small enough to allow the tunneling of electrons through the gap, charge transfer plasmonic modes emerge. Let us examine the spectra at the extreme limit of $d_{gap} = 0.1\text{nm}$ shown in Fig. S4(a) at which tunneling can surely occur. We see that in addition to the mode Q and mode M, two new plasmonic modes appear, denoted as *C1* and *C2*, respectively. While both are charge transfer modes, they have distinct properties. Examination of the current movement (Fig. S4(e)) suggests that the induced currents flow through the gap in mode C1 and the two particles effectively become a long dumbbell. This explains the low frequency of the mode. The mode C2 originates from the mode D as can be seen in Fig. S4(a). Comparing Figs. S4(b) and S4(d), we see that the induced charge distributions are similar except near the gap region, which means the C2 mode can be treated as a charge transfer corrected D mode.

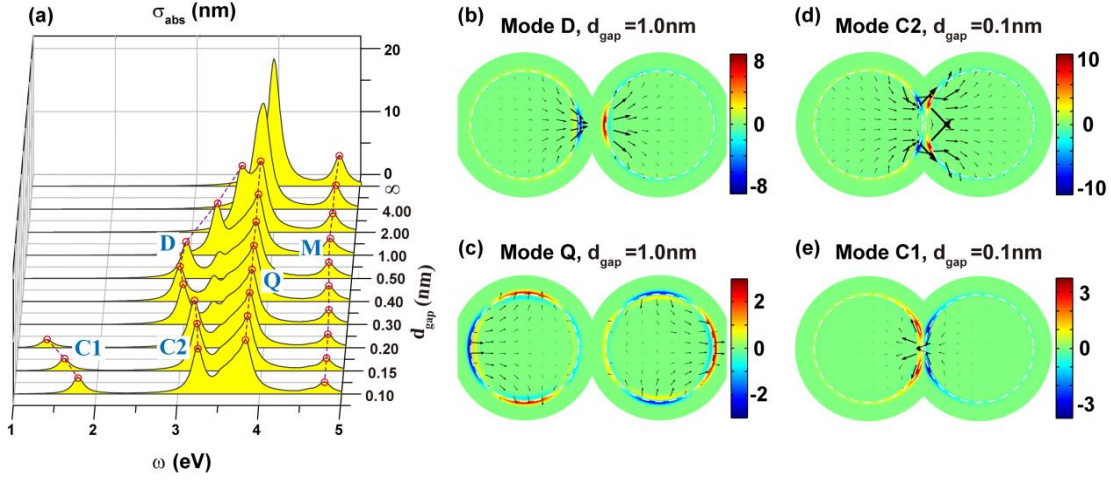

**Figure S4.** (a) Absorption cross section $\sigma_{abs}$ spectrum for different separations ($d_{gap}$) of the cylinder dimers under plane wave illumination with $\mathbf{k}_{inc} \parallel y$, $\mathbf{E}_{inc} \parallel x$ and $|\mathbf{E}_{inc}| = 1.0$ V/m. Dashed lines and open circles label different plasmonic modes. (b)-(e) Contour plots of the induced charges $\rho_1/E_0$ and arrow plot of the current densities $\mathbf{J}_1$ at some particular time for (b) D mode and (c) Q mode with $d_{gap} = 1.0$ nm; and for (d) C2 mode and (e) C1 mode with $d_{gap} = 0.1$ nm.

## III. Optical torque of bow-tie structures

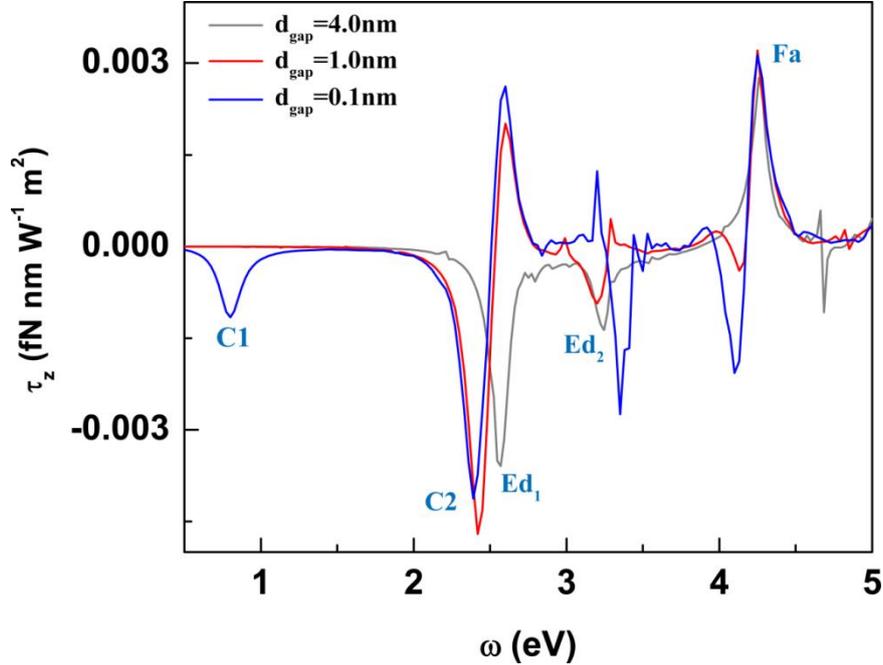

**Figure S5.** Optical torque (z-component) of the right-sided triangle in the two-triangle case under plane wave illumination. The solid gray, red, and blue lines are for the gap size 4.0nm, 1.0nm, and 0.1nm, respectively.